\documentclass[aps,prd,amsmath,eqsecnum,showpacs]{revtex4}
\usepackage{bm}
\usepackage{graphicx}
\begin{document}
\title{Dissipation in equations of motion of scalar fields}
\author{Ian D. Lawrie}
\email{i.d.lawrie@leeds.ac.uk}
\affiliation{Department of Physics and Astronomy, The University of Leeds, Leeds LS2 9JT, England.}
\date{\today}
\begin{abstract}
The methods of non-equilibrium quantum field theory are used to investigate the possibility
of representing dissipation in the equation of motion for the expectation value of a scalar
field by a friction term, such as is commonly included in phenomenological inflaton equations
of motion.  A sequence of approximations is exhibited which reduces the non-equilibrium theory
to a set of local evolution equations.  However, the adiabatic solution to these evolution
equations which is needed to obtain a local equation of motion for the expectation value is
not well defined;  nor, therefore, is the friction coefficient.  Thus, a non-equilibrium
treatment is essential, even for a system that remains close to thermal equilibrium, and
the formalism developed here provides one means of achieving this numerically.
\end{abstract}
\maketitle
\section{Introduction\label{intro}}

Inflationary universe scenarios frequently refer to an inflaton equation of motion
of the form
\begin{equation}
\ddot{\phi} + 3H\dot{\phi} + \eta(\phi)\dot{\phi}+V_{\mathrm{eff}}(\phi)=0.
\label{localeom}
\end{equation}
The friction term $\eta(\phi)\dot\phi$ represents one mechanism through which the inflationary
universe might be reheated, converting the inflaton energy into particles
\cite{dolgov1982,abbott1982,albrecht1982}, though the process of parametric amplification
\cite{traschen1990,kofman1994,kofman1997} is in some cases a more important effect. Moreover,
it was pointed out long ago by Moss \cite{moss1985} that a frictional effect large enough to
maintain a thermal bath of particles during the inflationary era would have a significant
effect both on the inflaton evolution and on the primordial perturbation spectrum, alleviating
the fine-tuning problems that afflict many inflationary models.  More recently, this idea has
been extensively investigated by Berera and co-workers
\cite{berera1995,bererafang1995,berera2000}.

While some frictional effect is to be expected in any interacting field theory, it is
by no means clear at a fundamental level that this effect can properly be represented
by a local equation of motion such as (\ref{localeom}), and the purpose of the
present paper is to study this question in some detail. (A brief account of our findings
has been given in Ref.~\cite{lawrie2002}.)  To be fairly generic, we may consider a scalar
field $\Phi$, which couples to other scalar fields and to fermions,
denoted collectively by $\xi$ and $\psi$ respectively.  In Minkowski spacetime, a
schematic Lagrangian density for this collection of fields is
\begin{equation}
{\cal L} = {\textstyle\frac{1}{2}}\partial_\mu\Phi\partial^\mu\Phi
-{\textstyle\frac{1}{2}}m_\Phi^2\Phi^2
-{\textstyle\frac{1}{4!}}\lambda\Phi^4
-{\textstyle\frac{1}{2}}g_1\Phi^2\xi^2
-g_2\Phi\bar{\psi}\psi +\Delta{\cal L}(\xi,\psi)
\label{genlag}
\end{equation}
where $\Delta{\cal L}$ is that part of the Lagrangian that does not depend on $\Phi$.
(The considerations developed here can probably be extended with little difficulty
to a theory in which $\Phi$ also couples to gauge fields.  They are essentially unchanged
in a Robertson-Walker spacetime, as we discuss in section~\ref{discuss}.)
On splitting $\Phi(t,\bm{x})$ into its expectation value $\phi(t)$, which we take to
be spatially homogeneous, and a fluctuation field $\varphi(t,\bm{x})$ with
$\langle\varphi(t,\bm{x})\rangle=0$, we find that the equation of motion
for $\phi(t)$ is
\begin{equation}
\ddot{\phi}+m_\Phi^2\phi+{\textstyle\frac{1}{6}}\lambda\phi^3
+{\textstyle\frac{1}{2}}\lambda\phi\langle\varphi^2\rangle
+g_1\phi\langle\xi^2\rangle
+g_2\langle\bar{\psi}\psi\rangle
+{\textstyle\frac{1}{6}}\lambda\langle\varphi^3\rangle=0
\label{geneom}
\end{equation}
while at tree level, the quantum fields $\varphi$, $\xi$ and $\psi$ acquire effective
masses
\begin{equation}
m_\varphi^2(t) = m_\Phi^2+{\textstyle\frac{1}{2}}\lambda\phi^2(t)\qquad
m_\xi^2(t) = m_\xi^2+g_1\phi^2(t)\qquad
m_\psi(t) = m_\psi + g_2\phi(t)
\label{masses}
\end{equation}
which depend on time through $\phi(t)$.  Dissipative effects in the equation
of motion (\ref{geneom}) can arise from the final term
$\frac{1}{6}\lambda\langle\varphi^3\rangle$, but we concentrate here on those
arising from the other expectation values, which are quadratic in the quantum
fields.  Generically, the equation of motion is
\begin{equation}
\ddot{\phi}(t)+m_\Phi^2\phi(t)
+g\langle\chi(t,\bm{x})^2\rangle\phi(t)+\cdots =0
\label{phieom}
\end{equation}
where $\chi$ may stand for $\varphi$, the fluctuating part of $\Phi$, for another
scalar field $\xi$ or for a fermionic field $\psi$, and has a $\phi(t)$-dependent
mass of the form (\ref{masses}).  In general, there will be a sum of such contributions
from the fields appearing in a specific model, but our considerations are substantially
model-independent.

Superficially, it is quite plausible that a system which remains fairly close to
equilibrium can be treated by using equilibrium statistical mechanics.  In
section~\ref{etd}, we review the relevant dissipative mechanisms and the
results for the friction coefficient $\eta(\phi)$ which have been obtained on
this basis; we also indicate why this approach might be questioned, when applied
to a system that is not maintained in exact thermal equilibrium.
Sections~\ref{lase}-\ref{renorm} develop a strategy for tackling the problem
in non-equilibrium scalar field theory, exhibiting the sequence of approximations
necessary to obtain a local equation of motion of the form (\ref{localeom}).  We find
that this local equation of motion contains coefficients whose values must
be found from the solution of auxiliary kinetic equations.  The resulting set
of local evolution equations is suitable for a numerical solution.  However,
we investigate in section~\ref{friction} whether an adiabatic treatment of these
evolution equations leads to a well defined friction coefficient $\eta(\phi)$.
We find that it does not, and section~\ref{numeric} exhibits numerical
evidence that non-equilibrium effects may be quantitatively quite significant.
Section~\ref{fermions} briefly discusses frictional effects due to fermions,
showing that the formal situation is quite similar to that developed in detail
for scalar fields.  Finally, our principal conclusions are summarized in
section~\ref{discuss}, where we also discuss the extent to which they depend
on the methods of approximation we have utilized.

\section{Mechanisms of energy transfer and dissipation\label{etd}}
Roughly speaking, we can identify two mechanisms through which energy may
be transferred between the classical field $\phi$ and the system of $\chi$
particles.  One is the creation of new particles, which we will refer to as
type-I.  The other, to which we refer as type-II, involves changing the energies
of particles that are already present.  In the following subsections, we review
the arguments which purport to derive from these two mechanisms friction terms in
the equation of motion for $\phi$.
\subsection{Particle creation\label{typeI}}
A crude argument given by Morikawa and Sasaki \cite{morikawa84} takes $\chi$ in
the first instance to be a free field, except that the interaction with $\phi$
leads to a time-dependent effective mass.  Thus, the energy of a single $\chi$
particle is $\omega_k(t)=\sqrt{k^2+m^2+g\phi^2(t)}$.  According to standard
arguments, this field can be represented as
\begin{equation}
\chi(t,\bm{x})=\int\frac{d^3k}{(2\pi)^3}\left[e^{i\bm{k\cdot x}}f_k(t;\hat{t})a_k(\hat{t})
+e^{-i\bm{k\cdot x}}f_k^*(t;\hat{t})a_k^\dag(\hat{t})\right]
\label{chiexpansion}
\end{equation}
where the mode functions $f_k(t;\hat{t})$ are solutions of the equation
\begin{equation}
\left[\partial^2_t+\omega_k^2(t)\right]f_k(t;\hat{t})=0.
\label{feqn}
\end{equation}
They satisfy the Wronskian condition
\begin{equation}
f_k(t;\hat{t})\dot{f}^*_k(t;\hat{t})-\dot{f}_k(t;\hat{t})f^*_k(t;\hat{t})=i,
\label{wronskian}
\end{equation}
where the overdot indicates differentiation with respect to the first argument, $t$.
Of the many complex functions that obey these two equations, we select a family,
parametrized by the reference time $\hat{t}$, by imposing the boundary
conditions
\begin{equation}
f_k(\hat{t};\hat{t})=\left[2\omega_k(\hat{t})\right]^{-1/2}\qquad
\dot{f}_k(\hat{t};\hat{t})=-i\left[\omega_k(\hat{t})/2\right]^{1/2}.
\label{fbcs}
\end{equation}
These mode functions can be expressed as
\begin{equation}
f_k(t;\hat{t})=\left[2\Omega_k(t;\hat{t})\right]^{-1/2}\exp\left[-i\int_{\hat{t}}^t\Omega_k(t';\hat{t})dt'\right]
\label{wkb}
\end{equation}
where the frequency $\Omega_k(t;\hat{t})$ is the solution of
\begin{equation}
\frac{1}{2}\frac{\ddot{\Omega}_k}{\Omega_k}-\frac{3}{4}\frac{\dot{\Omega}_k^2}{\Omega_k^2}+\Omega_k^2=\omega_k^2
\label{omegaeqn}
\end{equation}
subject to the boundary conditions $\Omega_k(\hat{t};\hat{t})=\omega_k(\hat{t})$ and
$\dot{\Omega}_k(\hat{t};\hat{t})=\ddot{\Omega}_k(\hat{t};\hat{t})=0$.  At times close to $\hat{t}$,
therefore, $f_k(t;\hat{t})$ is the positive-energy solution,
\begin{equation}
f_k(t;\hat{t})\approx\left[2\omega_k(\hat{t})\right]^{-1/2}\exp\left[-i\omega_k(\hat{t})(t-\hat{t})\right],
\label{approxf}
\end{equation}
but at other times the frequency $\Omega_k(t;\hat{t})$ does not necessarily correspond to a
single-particle energy.  Mode functions referred to different reference times are related by a
Bogoliubov transformation.  To be concrete, let $f_k^0(t)=f_k(t;0)$.  Then $f_k(t;\hat{t})$ is
a linear combination of $f_k^0(t)$ and $f_k^{0*}(t)$, say $f_k(t;\hat{t})=A(\hat{t})f_k^0(t)
+B(\hat{t})f_k^{0*}(t)$.  It is a simple exercise using (\ref{wronskian}) and (\ref{fbcs}) to find
the Bogoliubov coefficients $A(\hat{t})$ and $B(\hat{t})$ and thus the dependence on $\hat{t}$ of
$f_k(t;\hat{t})$, which is
\begin{equation}
\partial_{\hat{t}}f_k(t;\hat{t})=i\omega_k(\hat{t})f_k(t;\hat{t})
-\frac{1}{2}\frac{\dot{\omega}_k(\hat{t})}{\omega_k(\hat{t})}f_k^*(t;\hat{t}).
\label{dthatf}
\end{equation}

The creation and annihilation operators in (\ref{chiexpansion}) have the commutation relation
$[a_k(\hat{t}),a_{k'}^\dag(\hat{t})]=(2\pi)^3\delta(\bm{k}-\bm{k}')$.  Their dependence on the
reference time $\hat{t}$ follows from the fact that $\chi(t,\bm{x})$ itself is independent of
$\hat{t}$; using the fact that $f_k(t;\hat{t})$ depends only on $k=\vert\bm{k}\vert$, we find
\begin{equation}
\partial_{\hat{t}}a_k(\hat{t})=-i\omega_k(\hat{t})a_k(\hat{t})
+\frac{1}{2}\frac{\dot{\omega}_k(\hat{t})}{\omega_k(\hat{t})}a_{-k}^\dag(\hat{t}).
\label{dadt}
\end{equation}

With this formalism in hand, we evaluate the expectation value in (\ref{phieom}) by choosing the
reference time $\hat{t}$ to be $t$.  For a spatially homogeneous state, the result is
\begin{equation}
\langle\chi(t,\bm{x})^2)\rangle
=\int\frac{d^3k}{(2\pi)^32\omega_k(t)}\left[1+2n_k(t)+2\mathrm{Re}\,\nu_k(t)\right],
\label{chisqev}
\end{equation}
where the functions $n_k(t)$ and $\nu_k(t)$ are defined by
\begin{eqnarray}
\langle a_k^\dag(t)a_{k'}(t)\rangle&=&(2\pi)^3\delta(\bm{k}-\bm{k}')n_k(t)\label{nkcan}\\
\langle a_k(t)a_{k'}(t)\rangle&=&(2\pi)^3\delta(\bm{k}+\bm{k}')\nu_k(t).\label{nukcan}
\end{eqnarray}
The functions $n_k(t)$ can be interpreted as occupation numbers for the single-particle
modes $f_k(t;\hat{t})$ when $t$ is close to $\hat{t}$, while $\nu_k(t)$ measures the
off-diagonality of the density matrix in the representation defined by these modes. The time
dependence of these functions is easily established from (\ref{dadt}), and we find
\begin{eqnarray}
\partial_tn_k(t)&=&\frac{\dot{\omega}_k(t)}{\omega_k(t)}\mathrm{Re}\,\nu_k(t)\label{dndtI}\\
\partial_t\nu_k(t)&=&-2i\omega_k(t)\nu_k(t)+\frac{\dot{\omega}_k(t)}{2\omega_k(t)}
\left[1+2n_k(t)\right].\label{dnudtI}
\end{eqnarray}

These evolution equations are valid only in the approximation that $\chi(t,\bm{x})$ is a free
field.  It is argued in \cite{morikawa84} that interactions will give rise to an imaginary part of
the $\chi$ self-energy, which can be taken into account by replacing $\omega_k(t)$ with
$\omega_k(t)-i\Gamma_k(t)$.  With the further assumption that $\omega_k(t)$, $\Gamma_k(t)$ and
$n_k(t)$ change negligibly on a time scale of order $\Gamma_k(t)^{-1}$, integration of (\ref{dnudtI})
yields
\begin{equation}
\mathrm{Re}\,\nu_k(t)\approx\frac{\dot{\omega}_k(t)\Gamma_k(t)\omega_k(t)}{2\left[\omega_k^2(t)+\Gamma_k^2(t)
\right]^2}\left[1+2n_k(t)\right]
\label{renums}
\end{equation}
if $\dot{\Gamma}_k$ is negligible by comparison with $\dot{\omega}_k$.  Since
$\dot{\omega}_k=g\phi\dot{\phi}/\omega_k$, we identify a contribution of the form
$\eta_{\mathrm{I}}(\phi)\dot{\phi}$ in the equation of motion (\ref{phieom}), with
\begin{equation}
\eta_{\mathrm{I}}(\phi)=\frac{g^2\phi^2(t)}{2}
\int\frac{d^3k}{(2\pi)^3}\frac{\Gamma_k(t)}{\omega_k(t)[\omega_k^2(t)
+\Gamma_k^2(t)]^2}\left[1+2n_k(t)\right].\label{etaI}
\end{equation}

\subsection{Changing particle energies\label{typeII}}
To the extent that the $\chi$ particles can be considered as free, the energy of this system of
particles can be estimated as
\begin{equation}
E(t)\approx\int\frac{d^3k}{(2\pi)^3}\,\omega_k(t)n_k(t).
\end{equation}
Clearly, this energy is altered by a change in the classical field $\phi(t)$, which causes a change in
the single-particle energies $\omega_k(t)$, and this provides a mechanism whereby energy may be exchanged.
However, a frictional term in the equation of motion for $\phi$ arises only from an irreversible
transfer of energy.  This is a secondary effect, brought about by the fact that a change in $\omega_k(t)$
alters the scattering and decay rates of $\chi$ particles, and hence affects the evolution of the occupation
numbers $n_k(t)$.  A crude argument for estimating the friction coefficient that arises from this
mechanism was given by Hosoya and Sakagami \cite{hosoya84}, who take the time evolution of $n_k(t)$ to be
governed by a phenomenological kinetic equation of the form
\begin{equation}
\partial_tn_k(t)=-2\Gamma_k\left[n_k(t)-n^{\mathrm{eq}}_k(t)\right],\label{dndtII}
\end{equation}
which is sometimes referred to as the relaxation-time approximation to the Boltzmann equation.  It is
implied that the system of particles is always close to an equilibrium state characterized by a
fixed temperature $\beta^{-1}$, with the distribution function
\begin{equation}
n_k^{\mathrm{eq}}(t)=1/\left[\exp(\beta\omega_k(t))-1\right].
\label{bedist}
\end{equation}
If we once more assume that $n_k^{\mathrm{eq}}(t)$ changes slowly on a time scale $\Gamma_k^{-1}$, then
(\ref{dndtII}) can be approximately integrated to yield $n_k(t)=n_k^{\mathrm{eq}}(t)+\delta n_k(t)$, with
\begin{eqnarray}
\delta n_k(t)&\approx& -(2\Gamma_k)^{-1}\partial_tn^{\mathrm{eq}}(t)\nonumber\\
&\approx& \beta(2\Gamma_k)^{-1}\dot{\omega}_k(t)n^{\mathrm{eq}}(t)\left[
1+n^{\mathrm{eq}}(t)\right]\label{nkhs}
\end{eqnarray}
and we obtain a contribution to the equation of motion (\ref{phieom}) of the form
$\eta_{\mathrm{II}}(\phi)\dot{\phi}$, with
\begin{equation}
\eta_{\mathrm{II}}(\phi)=\frac{\beta g^2\phi^2(t)}{2}\int\frac{d^3k}{(2\pi)^3}
\frac{n^{\mathrm{eq}}(t)\left[1+n^{\mathrm{eq}}(t)\right]}{\Gamma_k\omega_k^2(t)}.
\label{etaII}
\end{equation}

\subsection{Linear response theory\label{lrt}}

The crude arguments given above can in some respects be improved by restricting attention to a
situation in which the system of $\chi$ particles is in thermal equilibrium, apart from a small
time-dependent perturbation which is treated to linear order.  Different versions of this treatment
have been given by Hosoya and Sakagami \cite{hosoya84}, by Morikawa and Sasaki
\cite{morikawa84,morikawa85} and more recently by Berera, Gleiser and Ramos
\cite{gleiser94,berera98,berera01}.  It is necessary to suppose that over a sufficiently long
period of time, the classical field $\phi(t)$ can be decomposed as $\phi(t)=\phi_0+\delta\phi(t)$,
where $\phi_0$ is constant and $\delta\phi(t)$ is small.  Correspondingly, the Hamiltonian for $\chi$
will be decomposed as
\begin{equation}
H(\chi,t)=H_0(\chi)+g\phi_0\delta\phi(t)\int d^3x\,\chi^2(t,\bm{x})+{\mathrm{O}}(\delta\phi^2),
\end{equation}
the second term being treated as a perturbation.  To linear order in this perturbation, the
standard Kubo formalism then yields
\begin{equation}
\langle\chi^2(t,\bm{x})\rangle\approx\langle\chi^2(t,\bm{x})\rangle^{\mathrm{eq}}
+ig\phi_0\int_{-\infty}^tdt'\delta\phi(t')\int d^3x'\langle\left[\chi^2(t',\bm{x}'),\chi^2(t,\bm{x})
\right]\rangle^{\mathrm{eq}}\label{linres}
\end{equation}
where $^{\mathrm{eq}}$ denotes the thermal average in the equilibrium ensemble determined by
$H_0(\chi)$.  Although $H_0(\chi)$ has no explicit time dependence, it still contains interactions,
and the remaining thermal averages can be computed systematically in perturbation theory.  To obtain
a well-defined answer for the linear response term in (\ref{linres}), it proves necessary to effect a
partial resummation of the $\chi$ propagator, in particular identifying the thermal width $\Gamma_k$
as the imaginary part of a suitable self-energy (as discussed in more detail below). As first pointed out
in Ref. \cite{morikawa85}, the net result is a contribution to the equation of motion (\ref{phieom})
of the form $\eta(\phi)\dot{\phi}$, with $\eta(\phi)\approx\eta_{\mathrm{I}}(\phi_0)
+\eta_{\mathrm{II}}(\phi_0)$, provided that $\Gamma_k\ll\omega_k$ and that $n_k(t)$ in (\ref{etaI})
and $n^{\mathrm{eq}}(t)$ in (\ref{etaII}) are identified as the constant equilibrium distribution associated
with $H_0(\chi)$.  As in the previous calculations, it is also necessary to assume that $\delta\phi(t')$
varies slowly on a time scale of order $\Gamma_k^{-1}$ and can be approximated as
$\delta\phi(t')\approx\dot{\phi}(t)(t'-t)$.

\subsection{Open questions}
Each of the calculations outlined above has its own deficiencies.  The type-I analysis of section
\ref{typeI} is valid arbitrarily far from equilibrium, but treats interactions in an incomplete
and \textit{ad hoc} manner.  The type-II analysis of section \ref{typeII} is explicitly restricted
to states very close to local equilibrium and rests on a kinetic equation which is little more
than a guess.  The linear response treatment of section \ref{lrt} is much more systematic, insofar
as the expectation value in (\ref{linres}) can in principle be evaluated at any desired order
of equilibrium perturbation theory.  However, this calculation depends in an essential way on
analytic properties of thermal Green functions that are meaningful only in a state of exact thermal
equilibrium.  Specifically, the Wightman functions
$G^{>}(t-t';\bm{x}-\bm{x}')=\langle\chi(t,\bm{x})\chi(t',\bm{x}')\rangle^{\mathrm{eq}}$ and
$G^{<}(t-t';\bm{x}-\bm{x}')=\langle\chi(t',\bm{x}')\chi(t,\bm{x})\rangle^{\mathrm{eq}}$ have Fourier
transforms which satisfy the KMS condition $G^{>}(k_0,\bm{k})=\exp(\beta k_0)G^{<}(k_0,\bm{k})$
(see, for example, Ref. \cite{lebellac00}).  In a state which departs even slightly from thermal
equilibrium, neither the temporal Fourier transform nor the KMS condition has any meaning.

The primary question addressed in the remainder of this paper is whether the apparently plausible
result of linear response theory is recovered, for a non-equilibrium system, in the limit of slow
time evolution.  This will be possible, at best, only if we have a local approximation to expectation
values such as $\langle\chi^2(t,\bm{x})\rangle$, which are inherently nonlocal in time (as evidenced by
(\ref{linres}) even in the linear response approximation), and devising such an approximation is the
key feature of the analysis that follows.

An important issue that is not addressed in this paper arises from the expression (\ref{etaII}).  This
contribution to the friction coefficient depends inversely on the thermal width $\Gamma_k$, which is typically
of the order of the square of a coupling constant, and appears to call into question the reliability of
perturbation theory as applied to this problem.  In fact, this is typical of expressions obtained in the
application of linear response theory to the estimation of transport coefficients, and it is known that
infinite classes of diagrams contribute at each order of perturbation theory \cite{jeon93,jeon95,jeon96}.
How the requisite resummation might be effected for a nonequilibrium system is beyond the scope of this
paper.

\section{Local approximation for self-energies\label{lase}}
\begin{figure}
\includegraphics{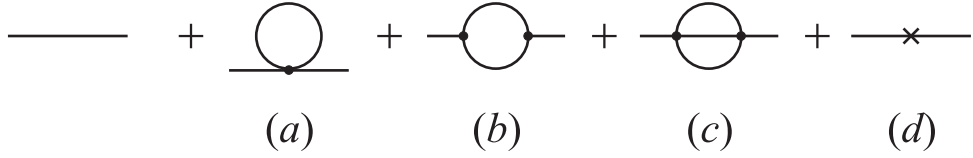}
\caption{\label{fig1}Diagrams contributing to the 2-point function of $\lambda\Phi^4$ theory.
Diagram ($a$) gives a local contribution to the self-energy; diagram ($c$) gives rise to the principal
dissipative effects; diagram ($d$) represents the contribution of ${\cal M}_{ab}$.}
\end{figure}

The exact two-point functions for the quantum field $\chi(t,\bm{x})$ will be denoted in a standard
notation by
\begin{equation}
G_{ab}(t,t';\bm{k})=\int d^3x\,e^{i\bm{k\cdot x}}G_{ab}(t,\bm{x};t',\bm{0}),
\label{GFT}
\end{equation}
(for $a,b = 1,2$) where
\begin{equation}
G(t,\bm{x};t',\bm{x}')=
\begin{pmatrix}
\langle T[\chi(t,\bm{x})\chi(t',\bm{x}')]\rangle &
\langle \chi(t',\bm{x}')\chi(t,\bm{x})\rangle\\
\langle \chi(t,\bm{x})\chi(t',\bm{x}')\rangle &
\langle \bar{T}[\chi(t,\bm{x})\chi(t',\bm{x}')]\rangle
\end{pmatrix}.
\label{Gdef}
\end{equation}
Self-energies $\Sigma_{ab}(t,t';\bm{k})$ can be defined by the Dyson-Schwinger equations
\begin{equation}
G_{ab}(t,t';\bm{k})=g_{ab}^{(\mathrm{F})}(t,t';\bm{k})-i\int dt''dt'''g^{(\mathrm{F})}_{ac}(t,t'';\bm{k})
\Sigma_{cd}(t'',t''';\bm{k})G_{db}(t''',t';\bm{k}),
\label{dseqn}
\end{equation}
in which the free-field propagators $g_{ab}^{(\mathrm{F})}(t,t';\bm{k})$ are solutions of the equations
\begin{equation}
{\cal D}_{ac}^{(\mathrm{F})}(t,\partial_t;\bm{k})g_{cb}^{(\mathrm{F})}(t,t';\bm{k})
=g_{ac}^{(\mathrm{F})}(t,t';\bm{k})\overleftarrow{\cal D}_{cb}^{(\mathrm{F})}(t',\partial_{t'};\bm{k})
=-i\delta_{ab}\delta(t-t'),
\label{ffpeqn}
\end{equation}
where the differential operator ${\cal D}^{(\mathrm{F})}(t,\partial_t;\bm{k})$ is given by
\begin{equation}
{\cal D}^{(\mathrm{F})}=
\begin{pmatrix}
\partial_t^2+k^2+m^2(t) & 0\\
0 & -\partial_t^2-k^2-m^2(t)
\end{pmatrix}
\end{equation}
with $m^2(t)=m^2+g\phi^2(t)$. The form of the self-energy matrix $\Sigma_{ab}$ is constrained by some
general considerations. First, the full propagators $G_{ab}(t,t';\bm{k})$ defined by (\ref{GFT}) and
(\ref{Gdef}) have the properties
\begin{equation}
G_{ab}(t,t';\bm{k})=G_{ba}(t',t;\bm{k})
\end{equation}
\begin{equation}
G^*_{11}(t,t';\bm{k})=G_{22}(t,t';\bm{k}),\qquad G^*_{12}(t,t';\bm{k})=G_{21}(t,t';\bm{k}).
\end{equation}
Second, causality requires the integrand in (\ref{dseqn}) to vanish if either $t''$ or $t'''$ is larger
than both $t$ and $t'$.  From these observations, it is not hard to show that the self-energy
matrix has the general form
\begin{equation}
\Sigma(t,t';\bm{k})=\Sigma^{\mathrm{L}}(t;\bm{k})\delta(t-t')+\Sigma^{>}(t,t';\bm{k})\theta(t-t')
+\Sigma^{<}(t,t';\bm{k})\theta(t'-t)
\end{equation}
with
\begin{eqnarray}
\Sigma^{\mathrm{L}}(t;\bm{k})&=&
\begin{pmatrix}
\beta_k(t)-i\alpha_k(t)&i\alpha_k(t)\\ i\alpha_k(t)&-\beta_k(t)-i\alpha_k(t)
\end{pmatrix}
\label{localse}\\
\Sigma^{>}(t,t';\bm{k})&=&
\begin{pmatrix}
A_k(t,t')&A^*_k(t,t')\\ -A_k(t,t')&-A^*_k(t,t')
\end{pmatrix}\\
\Sigma^{<}(t,t';\bm{k})&=&
\begin{pmatrix}
A_k(t',t)&-A_k(t',t)\\ A^*_k(t',t)&-A^*_k(t',t)
\end{pmatrix}.
\end{eqnarray}
In the local part $\Sigma^{\mathrm{L}}$, the functions $\alpha_k(t)$ and $\beta_k(t)$ are real.
If $\chi$ has, for example, a quartic self-coupling $\chi^4$, or a biquadratic coupling
$\chi^2\xi^2$ to another scalar field $\xi$, then these couplings generate Feynman diagrams
(for example, diagram $(a)$ in figure~\ref{fig1}) that
are manifestly local; such diagrams contribute to $\beta_k(t)$, but not to $\alpha_k(t)$.  More
generally, it will be advantageous to extract local contributions from non-local diagrams,
and it is easily seen that a contribution to $A_k(t,t')$ of the form
$[\beta_k(t)-i\alpha_k(t)]\delta(t-t')$ is equivalent to (\ref{localse}), provided that we identify
$\theta(0)=\frac{1}{2}$.  We take $\alpha_k(t)=0$ for now, but will later use $\alpha_k(t)$ to denote
a local contribution to the imaginary part of $A_k(t,t')$.

In fact, the goal of this section is to develop a plausible \textit{ansatz} for approximating the whole self-energy
matrix as a local quantity, containing terms proportional only to $\delta(t-t')$ and
$\partial_t\delta(t-t')$.  To this end, it is convenient to deal with the commutator and anticommutator
(or correlation) functions
\begin{eqnarray}
\rho(t,t';\bm{k})&=&i\int d^3x\,e^{i\bm{k\cdot x}}
\langle[\chi(t,\bm{x}),\chi(t',\bm{0})]\rangle\nonumber\\
&=&i\left[G_{21}(t,t';\bm{k})-G_{12}(t,t';\bm{k})\right]\\
C(t,t';\bm{k})&=&{\textstyle\frac{1}{2}}\int d^3x\,e^{i\bm{k\cdot x}}
\langle\{\chi(t,\bm{x}),\chi(t',\bm{0})\}\rangle\nonumber\\
&=&{\textstyle\frac{1}{2}}\left[G_{21}(t,t';\bm{k})+G_{12}(t,t';\bm{k})\right]
\end{eqnarray}
from which all the $G_{ab}$ can be constructed. For these functions, the Dyson-Schwinger
equations (\ref{dseqn}) together with (\ref{ffpeqn}) imply the equations of motion (see, for example,
Ref. \cite{aarts})
\begin{eqnarray}
[\partial_t^2+\omega_k^2(t)]\rho(t,t';\bm{k})&=&-\int_{t'}^t dt''\,\Sigma_\rho(t,t'';\bm{k})\rho(t'',t';\bm{k})
\label{rhoeqn}\\
{[}\partial_t^2+\omega_k^2(t)]C(t,t';\bm{k})&=&-\int_0^t dt''\,\Sigma_\rho(t,t'';\bm{k})C(t'',t';\bm{k})
+\int_0^{t'}dt''\,\Sigma_C(t,t'';\bm{k})\rho(t'',t;\bm{k}),\nonumber\\\label{ceqn}
\end{eqnarray}
with
\begin{eqnarray}
\omega_k^2(t)&=&k^2+m^2(t)+\beta_k(t)\label{omegakdef}\\
\Sigma_\rho(t,t';\bm{k})&=&\left[A_k(t,t')+A^*_k(t,t')\right]\theta(t-t')
-\left[A_k(t',t)+A_k^*(t',t)\right]\theta(t'-t)\label{sigmarhodef}\\
\Sigma_C(t,t';\bm{k})&=&-{\textstyle\frac{1}{2}}i\left[A_k(t,t')-A^*_k(t,t')\right]\theta(t-t')
-{\textstyle\frac{1}{2}}i\left[A_k(t',t)-A_k^*(t',t)\right]\theta(t'-t).\nonumber\\
\label{sigmacdef}
\end{eqnarray}

In a state of thermal equilibrium, for which $\rho(t,t';\bm{k})=\rho^{\mathrm{eq}}(t-t';\bm{k})$, one can define
the spectral density
\begin{equation}
\rho^{\mathrm{eq}}(\omega,\bm{k})=-i\int dt\,e^{i\omega t}\rho^{\mathrm{eq}}(t;\bm{k}).
\end{equation}
A strategy frequently adopted in the context of linear response theory is to assume that this spectral
density can be approximated by the Breit-Wigner form
\begin{equation}
\rho^{\mathrm{eq}}(\omega,\bm{k})\approx\frac{4\Gamma_k\omega}{(\omega^2-\Omega_k^2-\Gamma_k^2)^2+4\Gamma_k^2\omega^2}
\end{equation}
where $\Omega_k$ and $\Gamma_k$ are a quasiparticle energy and width to be abstracted from the real and
imaginary parts of the self-energies. The numerical calculations of Ref.~\cite{aarts} for a one-dimensional
scalar field theory suggest that such an approximation is reasonable (though certainly not exact) for
a non-equilibrium state also.  To realize this approximation in the present context, observe that the real-time
commutator function is given by the Breit-Wigner approximation as
\begin{equation}
\rho^{\mathrm{eq}}(t-t';\bm{k})\approx \frac{i}{2\Omega_k}\left[e^{-i\Omega_k(t-t')}-e^{i\Omega_k(t-t')}\right]
e^{-\Gamma_k\vert t-t'\vert}.
\end{equation}
For the non-equilibrium state, we adopt an \textit{ansatz} which is a natural generalization of this expression,
namely $\rho(t,t';\bm{k})\approx\rho^{(0)}(t,t';\bm{k})$, where
\begin{eqnarray}
\rho^{(0)}(t,t';\bm{k})&=&i\left[f_k(t;\hat{t})f^*_k(t';\hat{t})-f^*_k(t;\hat{t})f_k(t',\hat{t})\right]
\left[\Theta_k(t,t')+\Theta_k(t',t)\right]\label{rho0}\\
\Theta_k(t,t')&=&\exp\left[-\int_{t'}^t dt''\Gamma_k(t'')\right]\theta(t-t')\label{thetadef}
\end{eqnarray}
and $f_k(t;\hat{t})$ are the mode functions introduced in section \ref{typeI}, except that the quasiparticle
energies $\omega_k(t)$ may include loop corrections. It is easily checked that $\rho^{(0)}(t,t';\bm{k})$ is
real and antisymmetric in its time arguments, as it should be.  It also has the property
$\left.\partial_t\rho^{(0)}(t,t';\bm{k})\right\vert_{t'=t}=1$ as required by the canonical commutator.
Finally, as can be verified from (\ref{dthatf}), it is independent of the reference time $\hat{t}$, provided
that $\Gamma_k(t)$ is. The approximate version of the equation of motion (\ref{rhoeqn}) obeyed by
$\rho^{(0)}(t,t';\bm{k})$ is conveniently expressed by writing
\begin{equation}
\rho^{(0)}(t,t';\bm{k})=\rho^{(0)>}(t,t';\bm{k})\theta(t-t')+\rho^{(0)<}(t,t';\bm{k})\theta(t'-t),
\end{equation}
with $\rho^{(0)<}(t,t';\bm{k})=-\rho^{(0)>}(t',t;\bm{k})$.  By differentiating (\ref{rho0}, we find
\begin{subequations}
\label{approxrhoeqn}
\begin{eqnarray}
\left[\partial_t^2+2\Gamma_k(t)\partial_t+\omega_k^2(t)+\Gamma_k^2(t)
+\dot{\Gamma}_k(t)\right]\rho^{(0)>}(t,t';\bm{k})&=&0\\
\left[\partial_t^2-2\Gamma_k(t)\partial_t+\omega_k^2(t)+\Gamma_k^2(t)
-\dot{\Gamma}_k(t)\right]\rho^{(0)<}(t,t';\bm{k})&=&0,
\end{eqnarray}
\end{subequations}
together with the equal-time conditions
\begin{equation}
\left.\partial_t\rho^{(0)>}(t,t';\bm{k})\right\vert_{t'=t}
=\left.\partial_t\rho^{(0)<}(t,t';\bm{k})\right\vert_{t'=t}=1.
\label{eqtimerho}
\end{equation}
Comparing these equations with (\ref{rhoeqn}), we see that they imply a corresponding
\textit{ansatz} for the self-energy (\ref{sigmarhodef}), or for the real part
of $A_k(t,t')$.  It is
\begin{equation}
\mathrm{Re}\,A_k^{(0)}(t,t')=\left[\Gamma_k(t)\Gamma_k(t')\right]^{1/2}\partial_t\delta(t-t'-\epsilon),
\label{approxsigmarho}
\end{equation}
where the positive infinitesimal $\epsilon$ is included to ensure that the delta function is satisfied
inside the range of integration in (\ref{rhoeqn}).

Given a nonzero quasiparticle width $\Gamma_k(t)$, the non-local self-energies may be expected
to decay roughly exponentially with $\vert t-t'\vert$.  For $\Sigma_C(t,t';\bm{k})$,
which is symmetric in its time arguments, or, equivalently, for the imaginary part of $A_k(t,t')$,
a suitable local \textit{ansatz} is
\begin{equation}
\mathrm{Im}\,A_k^{(0)}(t,t')=-\alpha_k(t)\delta(t-t').
\label{approxsigmac}
\end{equation}
With this approximation, the equation of motion (\ref{ceqn}) for the correlation function
becomes
\begin{equation}
\left[\partial_t^2+2\Gamma_k(t)\partial_t+\omega_k^2(t)+\Gamma_k^2(t)
+\dot{\Gamma}_k(t)\right]C^{(0)}(t,t';\bm{k})=-\alpha_k(t)\rho^{(0)}(t,t';\bm{k})\theta(t'-t).
\end{equation}
Alternatively, defining
\begin{equation}
C^{(0)}(t,t';\bm{k})=C^{(0)>}(t,t';\bm{k})\theta(t-t')+C^{(0)<}(t,t';\bm{k})\theta(t'-t),
\end{equation}
with $C^{(0)<}(t,t';\bm{k})=C^{(0)>}(t',t;\bm{k})$, this may be written as
\begin{subequations}
\label{approxceqn}
\begin{eqnarray}
\left[\partial_t^2+2\Gamma_k(t)\partial_t+\omega_k^2(t)+\Gamma_k^2(t)
+\dot{\Gamma}_k(t)\right]C^{(0)>}(t,t';\bm{k})&=&0\\
\left[\partial_t^2+2\Gamma_k(t)\partial_t+\omega_k^2(t)+\Gamma_k^2(t)
+\dot{\Gamma}_k(t)\right]C^{(0)<}(t,t';\bm{k})&=&-\alpha_k(t)\rho^{(0)<}(t,t';\bm{k})
\end{eqnarray}
\end{subequations}
together with the equal-time condition
\begin{equation}
\left.\partial_tC^{(0)>}(t,t';\bm{k})\right\vert_{t'=t}=\left.\partial_tC^{(0)<}(t,t';\bm{k})\right\vert_{t'=t}.
\label{eqtimec}
\end{equation}

Finally, we assemble the approximate correlation and commutator functions into a single complex
function,
\begin{subequations}
\begin{eqnarray}
C^{(0)>}(t,t',\bm{k})-{\textstyle\frac{1}{2}}i\rho^{(0)>}(t,t';\bm{k})&=&h_k(t,t')\\
C^{(0)<}(t,t',\bm{k})+{\textstyle\frac{1}{2}}i\rho^{(0)<}(t,t';\bm{k})&=&h_k(t',t).
\end{eqnarray}
\end{subequations}
Then, denoting by $g_{ab}(t,t';\bm{k})$ the approximation to the propagator matrix (\ref{GFT}) that
embodies the \textit{ans\"atze} (\ref{approxsigmarho}) and (\ref{approxsigmac}), this matrix can be
summarized by
\begin{equation}
g_{ab}(t,t';\bm{k})=h_b(t,t';\bm{k})\theta(t-t')+h_a(t',t;\bm{k})\theta(t'-t),
\label{approxgab}
\end{equation}
with $h_1(t,t';\bm{k})=h_k(t,t')$ and $h_2(t,t';\bm{k})=h_k^*(t,t')$. The approximate equations
of motion (\ref{approxrhoeqn}) and (\ref{approxceqn}) are
\begin{equation}
\left[\partial_t^2+2\Gamma_k(t)\partial_t+\omega_k^2(t)+\Gamma_k^2(t)
+\dot{\Gamma}_k(t)\right]h_k(t,t')=0\label{leftheqn}
\end{equation}
\begin{equation}
\left[\partial_t^2+\omega_k^2(t)+\Gamma_k^2(t)
-i\alpha_k(t)\right]h_k(t',t)+
\left[2\Gamma_k(t)\partial_t+\dot{\Gamma}_k(t)+i\alpha_k(t)\right]h_k^*(t',t)=0
\label{rightheqn}
\end{equation}
while the equal-time conditions (\ref{eqtimerho}) and (\ref{eqtimec}) become
\begin{equation}
\left.\partial_t\left[h_k(t,t')-h_k(t',t)\right]\right\vert_{t'=t}=-i\qquad
\left.\partial_t\left[h_k^*(t,t')-h_k(t',t)\right]\right\vert_{t'=t}=0.
\label{equaltimes}
\end{equation}

At this point, the \textit{ans\"atze} (\ref{approxsigmarho}) and (\ref{approxsigmac}) have yielded
local equations of motion for the approximate two-point functions, but the functions $\alpha_k(t)$
and $\Gamma_k(t)$ that appear in these equations and the function $\beta_k(t)$ that appears in the
single-particle energy (\ref{omegakdef}) are unknown.  Given a specific Lagrangian, approximations
to the self-energies $\Sigma_{ab}$ can be obtained---for example, from some version of perturbation
theory.  The function $\beta_k(t)$ can be identified from the local part of $\Sigma_{ab}$, but a
prescription is needed for extracting from the calculated $\Sigma_\rho$ and $\Sigma_C$ local
contributions of the kind indicated in (\ref{approxsigmarho}) and (\ref{approxsigmac}) so as to
identify $\Gamma_k(t)$ and $\alpha_k(t)$. This issue will be addressed in section \ref{renorm}.

\section{Local kinetic equations\label{kineq}}
The approximation summarized by (\ref{approxgab}) - (\ref{equaltimes}) is equivalent to that
obtained from somewhat different considerations in Ref.~\cite{lawrie89}.  There it was shown that the
general solution to the local equations of motion can be expressed in the form
\begin{eqnarray}
h_k(t,t')&=&{\textstyle\frac{1}{2}}\exp\left[-\int_{t'}^t dt''\,\Gamma_k(t'')\right]\tilde{h}_k(t,t')
\label{hdef}\\
\tilde{h}_k(t,t')&=&\left[1+Q_k(t';\hat{t})\right]
f_k(t;\hat{t})f_k^*(t';\hat{t})
+\left[-1+Q_k^*(t';\hat{t})\right]f_k^*(t;\hat{t})f_k(t';\hat{t}).
\end{eqnarray}
The function $Q_k(t;\hat{t})$ is a solution of the equation
\begin{equation}
\left[\partial_t+2\Gamma_k(t)+2\frac{\dot{f}_k^*(t;\hat{t})}{f_k^*(t;\hat{t})}\right]
\left[\partial_t+2\Gamma_k(t)\right]Q_k(t;\hat{t})=2i\alpha_k(t)
\label{qeqn}
\end{equation}
subject to the constraint
\begin{equation}
(\partial_t+2\Gamma_k)(Q_k+Q_k^*)+i(f_kf^*_k)^{-1}(Q_k-Q_k^*)=0,
\label{constraint}
\end{equation}
which is preserved by (\ref{qeqn}).  For our present purpose, it is useful to observe that
$Q_k(t;\hat{t})$ can be decomposed as
\begin{equation}
Q_k(t;\hat{t})=Q_k^{(1)}(t;\hat{t})+\frac{f_k(t;\hat{t})}{f_k^*(t;\hat{t})}Q_k^{(2)}(t;\hat{t})
\label{qdecomp}
\end{equation}
where $Q^{(1)}_k(t;\hat{t})$ is real.  If $Q_k^{(1)}$ and $Q_k^{(2)}$ are taken to obey the
first-order equations
\begin{eqnarray}
\left[\partial_t+2\Gamma_k(t)\right]Q_k^{(1)}(t;\hat{t})&=&2f_k(t;\hat{t})f_k^*(t;\hat{t})\alpha_k(t)
\label{q1eqn}\\
\left[\partial_t+2\Gamma_k(t)\right]Q_k^{(2)}(t;\hat{t})&=&-2f_k^{*2}(t;\hat{t})\alpha_k(t),
\label{q2eqn}
\end{eqnarray}
then both (\ref{qeqn}) and (\ref{constraint})
are satisfied and the decomposition (\ref{qdecomp}) is unique.

The propagator function $\tilde{h}_k(t,t')$ is now given by
\begin{eqnarray}
\tilde{h}_k(t,t')&=&\left[1+Q_k^{(1)}(t';\hat{t})\right]
f_k(t;\hat{t})f_k^*(t';\hat{t})
+\left[-1+Q_k^{(1)}(t';\hat{t})\right]f_k^*(t;\hat{t})f_k(t';\hat{t})\nonumber\\
&&\vphantom{2}+Q_k^{(2)}(t';\hat{t})f_k(t;\hat{t})f_k(t';\hat{t})
+Q_k^{(2)*}(t';\hat{t})f_k^*(t;\hat{t})f_k^*(t';\hat{t}).
\end{eqnarray}
It must be independent of the reference time $\hat{t}$, and
this determines the dependence on $\hat{t}$ of the functions $Q^{(i)}_k(t;\hat{t})$.  We find
\begin{eqnarray}
\partial_{\hat{t}}Q_k^{(1)}(t;\hat{t})&=&\frac{\dot{\omega}_k(\hat{t})}{\omega_k(\hat{t})}
\mathrm{Re}\,Q_k^{(2)}(t;\hat{t})\label{dhatq1}\\
\partial_{\hat{t}}Q_k^{(2)}(t;\hat{t})&=&-2i\omega_k(\hat{t})Q_k^{(2)}(t;\hat{t})
+\frac{\dot{\omega}_k(\hat{t})}{\omega_k(\hat{t})}Q_k^{(1)}(t;\hat{t}).\label{dhatq2}
\end{eqnarray}
In the equation of motion (\ref{phieom}), our approximation to $\langle\chi^2(t,\bm{x})\rangle$
is
\begin{equation}
\langle\chi^2(t,\bm{x})\rangle\approx\int\frac{d^3k}{(2\pi)^3}h_k(t,t)
\label{approxev}
\end{equation}
and this is conveniently evaluated by choosing the reference time $\hat{t}$ to be the time $t$ of
interest:
\begin{equation}
h_k(t,t)=Q_k^{(1)}(t;t)+\mathrm{Re}\,Q_k^{(2)}(t;t).
\end{equation}
In fact, let us define
\begin{eqnarray}
n_k(t)&=&{\textstyle\frac{1}{2}}\left[Q_k^{(1)}(t;t)-1\right]\label{nkdef}\\
\nu_k(t)&=&{\textstyle\frac{1}{2}}Q_k^{(2)}(t;t).\label{nukdef}
\end{eqnarray}
We have
\begin{equation}
h_k(t,t)=\left[2\omega_k(t)\right]^{-1}\left[1+2n_k(t)+2{\mathrm{Re}}\,\nu_k(t)\right]
\end{equation}
and will loosely identify the functions $n_k(t)$ and $\nu_k(t)$ with the quantities denoted by the
same symbols in section \ref{etd}.  By combining (\ref{q1eqn}) and (\ref{q2eqn}) with (\ref{dhatq1})
and (\ref{dhatq2}), we obtain the evolution equations
\begin{eqnarray}
\partial_tn_k(t)&=&\frac{\alpha_k(t)}{2\omega_k(t)}-\Gamma_k(t)\left[1+2n_k(t)\right]
+\frac{\dot{\omega}_k(t)}{\omega_k(t)}{\mathrm{Re}}\,\nu_k(t)\label{dndt}\\
\partial_t\nu_k(t)&=&-2i\left[\omega_k(t)-i\Gamma_k(t)\right]\nu_k(t)-\frac{\alpha_k(t)}{2\omega_k(t)}
+\frac{\dot{\omega}_k(t)}{2\omega_k(t)}\left[1+2n_k(t)\right].\label{dnudt}
\end{eqnarray}
These are clearly generalizations of the free-field equations (\ref{dndtI}) and (\ref{dnudtI}), the
extra terms involving the functions $\alpha_k(t)$ and $\Gamma_k(t)$ whose exact meanings are explored
further below.  On the other hand, the tentative identification
\begin{equation}
\alpha_k(t)\approx 2\omega_k(t)\Gamma_k(t)\left[1 + 2n^{\mathrm{eq}}_k(t)\right]
\label{naivealpha}
\end{equation}
(which we will later find to be too na\"{\i}ve) brings (\ref{dndt}) into the form
\begin{equation}
\partial_tn_k(t)\approx-2\Gamma_k(t)\left[n_k(t)-n^{\mathrm{eq}}_k(t)\right]
+\frac{\dot{\omega}_k(t)}{\omega_k(t)}{\mathrm{Re}}\,\nu_k(t),
\end{equation}
which is a generalization of the kinetic equation (\ref{dndtII}) including a source term to
account for particle creation.

\section{Determination of local self-energies\label{renorm}}
To give substance to the kinetic equations (\ref{dndt}) and (\ref{dnudt}), we need a concrete
method of determining the functions $\beta_k(t)$ (which appears in the quasiparticle energy
(\ref{omegakdef})), $\alpha_k(t)$ and $\Gamma_k(t)$. A prescription for doing this was given in
\cite{lawrie89}; here we describe a refinement of that prescription which is convenient for the
problem at hand.  The approximate two-point functions (\ref{approxgab}) which solve (\ref{leftheqn})
and (\ref{rightheqn}) are the exact propagators of an approximate theory defined by the closed-time-path
action
\begin{equation}
S^{(0)}_{\mathrm{CTP}}(\chi_1,\chi_2)=-{\textstyle\frac{1}{2}}\int dt\int d^3x\,\chi_a(t,\bm{x})
{\cal D}_{ab}(t,\partial_t,\bm{\nabla})\chi_b(t,\bm{x}),
\end{equation}
where the differential operator ${\cal D}$ is given, after a spatial Fourier transformation, by
\begin{equation}
{\cal D}=
\begin{pmatrix}
\partial_t^2+\omega_k^2(t)+\Gamma_k^2(t)-i\alpha_k(t) &
2\Gamma_k(t)\partial_t +\dot{\Gamma}_k(t)+i\alpha_k(t)\\
-2\Gamma_k(t)\partial_t -\dot{\Gamma}_k(t)+i\alpha_k(t) &
-\partial_t^2-\omega_k^2(t)-\Gamma_k^2(t)-i\alpha_k(t)
\end{pmatrix}.
\end{equation}
(An effective action having essentially this structure also describes an open system, coupled to
environmental degrees of freedom, which can be integrated out by the Feynman-Vernon influence
functional method \cite{hu1994,calzetta1994,hu1995}.  Here, one may think of a single field mode
having an environment that consists of all the other modes, but this environment is treated in a
self-consistent manner, rather than being integrated out.)
If the complete theory has the action $S(\chi)$, and the corresponding closed-time-path action
$S_{\mathrm{CTP}}(\chi_1,\chi_2) = S(\chi_1)-S(\chi_2)$, then a partly resummed perturbation
expansion can be defined by writing
\begin{equation}
S_{\mathrm{CTP}}(\chi_1,\chi_2)=S^{(0)}_{\mathrm{CTP}}(\chi_1,\chi_2)+\Delta S_{\mathrm{CTP}}(\chi_1,\chi_2)
\label{sctpsplit}
\end{equation}
and treating $\Delta S_{\mathrm{CTP}}$ as the perturbation.  Included in $\Delta S_{\mathrm{CTP}}$ is the counterterm
$\frac{1}{2}\int dt\,d^3x\,\chi_a{\cal M}_{ab}\chi_b$, with
\begin{eqnarray}
{\cal M}(t,\partial_t;\bm{k})&=&{\cal D}(t,\partial_t;\bm{k})-{\cal D}^{(\mathrm{F})}(t,\partial_t;\bm{k})
\nonumber\\
&=&\begin{pmatrix}\beta_k(t)+\Gamma_k^2(t)-i\alpha_k(t) &
2\Gamma_k(t)\partial_t +\dot{\Gamma}_k(t)+i\alpha_k(t)\\
-2\Gamma_k(t)\partial_t -\dot{\Gamma}_k(t)+i\alpha_k(t) &
-\beta_k(t)-\Gamma_k^2(t)-i\alpha_k(t)
\end{pmatrix},
\label{mabdef}
\end{eqnarray}
which accounts for the difference between $g_{ab}$ and $g^{(\mathrm{F})}_{ab}$.
In the context of this expansion, self-energies are defined by replacing $g^{(\mathrm{F})}_{ab}$ in the
Dyson-Schwinger equation (\ref{dseqn}) with $g_{ab}$. They have the form
\begin{equation}
\Sigma_{ab}(t,t';\bm{k})=\Sigma^{\mathrm{loop}}_{ab}(t,t';\bm{k})-{\cal M}_{ab}(t,\partial_t;\bm{k})\delta(t-t'),
\label{newse}
\end{equation}
where $\Sigma^{\mathrm{loop}}$ consists of loop diagrams in which the propagators are $g_{ab}$.

We would like to choose $\alpha_k(t)$, $\beta_k(t)$ and $\Gamma_k(t)$ in such a way that the propagators
$g_{ab}(t,t';\bm{k})$ approximate the exact two-point functions as closely as possible.  Loosely speaking,
this means making the self-energies (\ref{newse}) as small as possible.  More precisely, it is necessary
to obtain a local approximation to $\Sigma_{ab}^{\mathrm{loop}}$, which can be cancelled by an appropriate
choice of ${\cal M}_{ab}$.  To this end, suppose that $t$ and $t'$ are both close to the reference time
$\hat{t}$.  We define time-translation-invariant propagators which approximate $g_{ab}(t,t';\bm{k})$ in
this region by introducing the function
\begin{equation}
\bar{h}_k(\hat{t},\tau)=\frac{e^{-\hat{\Gamma}_k\tau}}{2\hat{\omega}_k}\left\{
[1+n_k(\hat{t})+\nu_k(\hat{t})]e^{-i\hat{\omega}_k\tau}+[n_k(\hat{t})+\nu_k^*(\hat{t})]
e^{i\hat{\omega}_k\tau}\right\},
\label{hbardef}
\end{equation}
where $\tau=t-t'$, $\hat{\omega}_k=\omega_k(\hat{t})$ and $\hat{\Gamma}_k=\Gamma_k(\hat{t})$. This function is obtained
from the one defined in (\ref{hdef}) by taking $\bar{h}_k(\hat{t},\tau)=h_k(\hat{t}+\tau,\hat{t})$, using the
approximation (\ref{approxf}) for $f_k(t;\hat{t})$, and replacing $\Gamma_k(t'')$ with $\hat{\Gamma}_k$ in
the prefactor.  The approximate propagators are then given by
\begin{equation}
\bar{g}_{ab}(\hat{t},\tau;\bm{k})=\bar{h}_b(\hat{t},\tau;\bm{k})\theta(\tau)
+\bar{h}_a(\hat{t},-\tau;\bm{k})\theta(-\tau),
\label{gbardef}
\end{equation}
where, as in (\ref{approxgab}), we use the notation $\bar{h}_1(\hat{t},\tau;\bm{k})=\bar{h}_k(\hat{t},\tau)$ and
$\bar{h}_2(\hat{t},\tau;\bm{k})=\bar{h}_k^*(\hat{t},\tau)$.

By replacing $g_{ab}$ with $\bar{g}_{ab}$ in the diagrams that constitute $\Sigma^{\mathrm{loop}}_{ab}$, we
obtain a time-translation invariant approximation to these self-energies,
$\bar{\Sigma}^{\mathrm{loop}}_{ab}(\hat{t},\tau;\bm{k})$, valid when $t$ and $t'$ are both close to $\hat{t}$,
which can be used to determine the functions $\alpha_k(\hat{t})$, $\beta_k(\hat{t})$ and $\Gamma_k(\hat{t})$.
After a Fourier transform on $\tau$, our approximation to the right-hand side of (\ref{newse}) is
\begin{equation}
\int\frac{d\omega}{2\pi}\left[\bar{\Sigma}^{\mathrm{loop}}_{ab}(\hat{t};\omega,\bm{k})-
{\cal M}_{ab}(\hat{t},-i\omega;\bm{k})\right]e^{-i\omega\tau}.
\label{rhs}
\end{equation}
Although $\bar{\Sigma}^{\mathrm{loop}}(\hat{t},\tau;\bm{k})$ is time-translation invariant, it is not, in general,
a distribution concentrated at $\tau=0$.  In Fourier transformed language,
$\bar{\Sigma}^{\mathrm{loop}}(\hat{t}; \omega,\bm{k})$ is a non-linear function of $\omega$, while
${\cal M}_{ab}(\hat{t},-i\omega;\bm{k})$ (which results from replacing $\partial_t$ in (\ref{mabdef})
with $-i\omega$) is linear in $\omega$.  Therefore, the integrand in (\ref{rhs}) cannot be made to vanish
for all values of $\omega$.  A reasonable prescription for determining $\alpha_k(\hat{t})$, $\beta_k(\hat{t})$
and $\Gamma_k(\hat{t})$ is to demand that
\begin{equation}
{\cal M}_{ab}(\hat{t},\pm i\hat{\omega}_k;\bm{k})
= \bar{\Sigma}^{\mathrm{loop}}_{ab}(\hat{t};\mp\hat{\omega}_k,\bm{k}).
\label{mabpresc}
\end{equation}
To the extent that the considerations of section \ref{lase} are valid, this ensures that $g_{ab}(t,t';\bm{k})$
are the propagators for free quasiparticles whose energies and widths are approximately those determined by
the peaks of the true non-equilibrium spectral density.  Explicitly, the prescription implied by
(\ref{mabpresc}) is
\begin{eqnarray}
\alpha_k(\hat{t})&=&-\frac{i}{2}\left[\bar{\Sigma}^{\mathrm{loop}}_{21}(\hat{t};\hat{\omega}_k,\bm{k})
+\bar{\Sigma}^{\mathrm{loop}}_{21}(\hat{t};-\hat{\omega}_k,\bm{k})\right]\label{alphapresc}\\
\Gamma_k(\hat{t})&=&-\frac{i}{4\hat{\omega}_k}\left[\bar{\Sigma}^{\mathrm{loop}}_{21}(\hat{t};\hat{\omega}_k,\bm{k})
-\bar{\Sigma}^{\mathrm{loop}}_{21}(\hat{t};-\hat{\omega}_k,\bm{k})\right]\label{gammapresc}\\
\beta_k(\hat{t})&=&\mathrm{Re}\,\bar{\Sigma}^{\mathrm{loop}}_{11}(\hat{t};\hat{\omega}_k,\bm{k})
-\Gamma_k^2(\hat{t}).\label{betapresc}
\end{eqnarray}

\section{Friction terms in the equation of motion\label{friction}}

The nature and purpose of the approximations we have introduced so far can usefully be summarized
as follows. By partitioning the closed-time-path action as in (\ref{sctpsplit}), we obtain a reorganized
perturbation theory in which the unperturbed propagators are the $g_{ab}$ defined by (\ref{approxgab}) -
(\ref{equaltimes}).  Summed to all orders, this perturbation theory would (presumably) be equivalent to
the usual expansion based on the free-particle propagators $g_{ab}^{(\mathrm{F})}$, and this assertion
is essentially independent of how we choose the functions $\alpha_k(t)$, $\beta_k(t)$
and $\Gamma_k(t)$ that enter the definition of $g_{ab}$. However, the reorganized perturbation
theory cannot in practice be summed to all orders.  Its utility rests on the possibility of making
$g_{ab}$ a better approximation than $g_{ab}^{(\mathrm{F})}$ to the full two-point functions; in particular,
we wish to estimate the expectation value $\langle\chi^2(t,\bm{x})\rangle$ by retaining only the lowest-order
term as indicated in (\ref{approxev}).  To do this, we choose $\alpha_k(t)$, $\beta_k(t)$
and $\Gamma_k(t)$ to be local contributions to the true self-energies, which are thereby
resummed in the reorganized perturbation expansion.  For the purpose of extracting these local contributions
(and \textit{only} for this purpose) we introduced the time-translation-invariant  propagators
$\bar{g}_{ab}$ in (\ref{gbardef}), which enabled us to formulate the prescription recorded in (\ref{alphapresc})
- (\ref{betapresc}).  Of course, this prescription can be implemented only approximately, by evaluating
$\bar{\Sigma}^{\mathrm{loop}}$ to some finite order of perturbation theory.

In principle, we are now in a position to evaluate the right-hand sides of the evolution equations
(\ref{dndt}) and (\ref{dnudt}), to solve these equations for $n_k(t)$ and $\nu_k(t)$ and hence to
estimate the expectation value (\ref{approxev}) in which we are principally interested.  There is, however,
a practical difficulty.  It is that the self-energy on the right-hand side of (\ref{gammapresc}) is itself
a function of $\Gamma_k$, and this equation cannot be solved analytically to obtain a concrete expression
for $\Gamma_k$.  A numerical solution is feasible, and this is no doubt the best way of estimating the
time evolution, given a specific model.  Here, though, we wish to investigate the circumstances under which
dissipation might be represented by the frictional terms in the equation of motion exhibited in section
\ref{etd}.  To that end, we now introduce two further approximations. First, we take the limit $\Gamma_k\to 0$
in the propagators $\bar{g}_{ab}$ used to calculate $\bar{\Sigma}_{ab}$ in (\ref{alphapresc}) - (\ref{betapresc}).
This is reasonably well justified in a weakly coupled theory, where $\Gamma_k$ is of order $g^2$, say, and
$\bar{\Sigma}_{ab}$ itself contains an overall factor of $g^2$.  Second, we will also set $\nu_k=0$ in
$\bar{g}_{ab}$.  A justification for this step will appear below [see equation (\ref{nukeq})].

With these approximations, the propagators $\bar{g}_{ab}$ assume the form familiar from the equilibrium theory
(see, for example, Ref. \cite{lebellac00}). In particular, the temporal Fourier transform of $\bar{g}_{12}$ is
\begin{equation}
\bar{g}_{21}(\hat{t};\omega,\bm{k})
=\frac{\pi}{\hat{\omega}_k}\left\{\left[1+n_k(\hat{t})\right]\delta(\omega-\hat{\omega}_k)
+n_k(\hat{t})\delta(\omega+\hat{\omega}_k)\right\}.
\label{gbar21}
\end{equation}
As a standard example, we consider in what follows the $\lambda\Phi^4$ theory, whose Lagrangian density
consists of the first three terms of (\ref{genlag}), identifying $\chi$ as the fluctuation field
$\varphi$ and the coupling $g$ as $g=\lambda/2$. The 2-point functions
for $\chi$ contain, amongst others, the diagrams shown in Fig.~\ref{fig1}. At 2-loop order, with $\bar{g}_{21}$
given by (\ref{gbar21}), the only contribution to $\bar{\Sigma}_{21}$ comes from diagram ($c$). (The 1-loop
diagram ($b$) contains products of $\delta$ functions which cannot be simultaneously satisfied.  It would
give an off-shell contribution if we were to retain a non-zero width $\Gamma_k$ in $\bar{g}_{21}$).
Evaluating this diagram, we find from (\ref{alphapresc}) and (\ref{gammapresc})
\begin{eqnarray}
\alpha_k(\hat{t})&=&\frac{\lambda^2}{32(2\pi)^5}\int d^3k_1d^3k_2d^3k_3
\frac{\delta(\hat{\omega}_{k_1}+\hat{\omega}_{k_2}-\hat{\omega}_{k_3}-\hat{\omega}_{k})
\delta(\bm{k}_1+\bm{k}_2-\bm{k}_3-\bm{k})}{\hat{\omega}_{k_1}\hat{\omega}_{k_2}\hat{\omega}_{k_3}}
\nonumber\\
&&\times\left[(1+\hat{n}_{k_1})(1+\hat{n}_{k_2})\hat{n}_{k_3}+\hat{n}_{k_1}\hat{n}_{k_2}(1+\hat{n}_{k_3})\right]
\label{alpharesult}\\
\Gamma_k(\hat{t})&=&\frac{\lambda^2}{64(2\pi)^5}\int d^3k_1d^3k_2d^3k_3
\frac{\delta(\hat{\omega}_{k_1}+\hat{\omega}_{k_2}-\hat{\omega}_{k_3}-\hat{\omega}_{k})
\delta(\bm{k}_1+\bm{k}_2-\bm{k}_3-\bm{k})}{\hat{\omega}_k\hat{\omega}_{k_1}\hat{\omega}_{k_2}\hat{\omega}_{k_3}}
\nonumber\\
&&\times\left[(1+\hat{n}_{k_1})(1+\hat{n}_{k_2})\hat{n}_{k_3}-\hat{n}_{k_1}\hat{n}_{k_2}(1+\hat{n}_{k_3})\right]
\label{gammaresult}
\end{eqnarray}
where $\hat{n}_{k_i}=n_{k_i}(\hat{t})$. In particular, the quantity
\begin{eqnarray}
S_k(\hat{t})&=&\frac{\alpha_k(\hat{t})}{2\hat{\omega}_k}-\Gamma_k(\hat{t})\left[1+2n_k(\hat{t})\right]
\nonumber\\
&=&\frac{\lambda^2}{32(2\pi)^5}\int d^3k_1d^3k_2d^3k_3
\frac{\delta(\hat{\omega}_{k_1}+\hat{\omega}_{k_2}-\hat{\omega}_{k_3}-\hat{\omega}_{k})
\delta(\bm{k}_1+\bm{k}_2-\bm{k}_3-\bm{k})}{\hat{\omega}_k\hat{\omega}_{k_1}\hat{\omega}_{k_2}\hat{\omega}_{k_3}}
\nonumber\\
&&\times\left[\hat{n}_{k_1}\hat{n}_{k_2}(1+\hat{n}_{k_3})(1+\hat{n}_k)
-(1+\hat{n}_{k_1})(1+\hat{n}_{k_2})\hat{n}_{k_3}\hat{n}_k\right]
\end{eqnarray}
which appears in the kinetic equation (\ref{dndt}) is precisely the 2-particle elastic scattering integral that
ought to appear in a genuine Boltzmann equation. (More generally, since the $\lambda\Phi^4$ theory has no
conserved particle number, inelastic contributions should also be expected, and these will indeed appear
if we extend the evaluation of the self-energies beyond 2-loop order.)

With these explicit expressions for $\alpha_k(t)$ and $\Gamma_k(t)$, we now have concrete forms for the
evolution equations (\ref{dndt}) and (\ref{dnudt}).  These evolution equations are local in time, but
their solutions will be non-local.  To extract contributions to $\langle\chi^2(t,\bm{x})\rangle$ which are
proportional to $\dot{\phi}(t)$, we must resort to some adiabatic approximation of the kind considered in
section \ref{etd}.  To do this systematically, we rewrite the evolution equations as
\begin{eqnarray}
\epsilon\partial_tn_k(t)&=&\bar{\alpha}_k(t)-\Gamma_k(t)\left[1+2n_k(t)\right]
+\epsilon\frac{\dot{\omega}_k(t)}{\omega_k(t)}{\mathrm{Re}}\,\nu_k(t)\label{dndt2}\\
\epsilon\partial_t\nu_k(t)&=&-2i\left[\omega_k(t)-i\Gamma_k(t)\right]\nu_k(t)-\bar{\alpha}_k(t)
+\epsilon\frac{\dot{\omega}_k(t)}{2\omega_k(t)}\left[1+2n_k(t)\right].\label{dnudt2}
\end{eqnarray}
where $\bar{\alpha}_k(t)=\alpha_k(t)/2\omega_k(t)$ and we have introduced a formal expansion parameter
$\epsilon$ multiplying terms with time derivatives.  By expanding in powers of $\epsilon$ and finally
setting $\epsilon=1$, we generate expansions of $n_k(t)$ and $\nu_k(t)$ in powers of the time derivatives
of $\phi(t)$.  On substituting these expansions in the expression (\ref{approxev}) for
$\langle\chi^2(t,\bm{x})\rangle$, the next-to-leading terms, proportional to $\dot{\phi}$, yield an
estimate of the friction coefficient $\eta(\phi)$.

At leading order, (\ref{dndt2}) reduces to the equation $S_k(t)=0$ whose solution is well-known to be the
Bose-Einstein distribution $n_k(t)=n_k^{\mathrm{eq}}(t)$ for some temperature $\beta^{-1}$. (We do not
allow for a non-zero chemical potential, because the inelastic contributions to $S_k(t)$ expected at
higher orders of perturbation theory would constrain the equilibrium chemical potential to vanish.)
The corresponding solution of (\ref{dnudt2}) is
\begin{equation}
\nu_k^{\mathrm{eq}}(t) = i\bar{\alpha}_k^{\mathrm{eq}}/2[\omega_k(t)-i\Gamma_k^{\mathrm{eq}}(t)]
\label{nukeq}
\end{equation}
where $\bar{\alpha}^{\mathrm{eq}}$ and $\Gamma_k^{\mathrm{eq}}$ are obtained from (\ref{alpharesult}) and
(\ref{gammaresult}) by setting $n_k = n_k^{\mathrm{eq}}$.  We see that, near equilibrium, $\nu_k$ is
smaller than $n_k$ by a factor of order $\Gamma_k/\omega_k$, so setting $\nu_k=0$ in the
calculation of the self-energies should be a fair approximation.

At next-to-leading order, we set $n_k \approx n_k^{\mathrm{eq}} + \epsilon\delta n_k$ and
$\nu_k \approx \nu_k^{\mathrm{eq}} + \epsilon\delta\nu_k$.  The linear-response approximation to
$\eta(\phi)$ [the sum of (\ref{etaI}) and (\ref{etaII})] can be recovered only at the expense
of further approximations.  The first is to replace $\Gamma_k(n)$ with $\Gamma_k^{\mathrm{eq}}
=\Gamma_k(n^{\mathrm{eq}})$ and similarly for $\bar{\alpha}_k(n)$:  that is, to ignore the
fluctuations in these quantities brought about by fluctuations in $n_k(t)$.  The second (when
calculating $\partial_t\nu^{\mathrm{eq}}$) is to take the time dependence of
$\Gamma_k^{\mathrm{eq}}(t)$ to be $\Gamma_k^{\mathrm{eq}}={\mathrm{constant}}/\omega_k(t)$.
These approximations amount to replacing the self-energies on the right-hand sides of
(\ref{alphapresc}) and (\ref{gammapresc}) with constant equilibrium values.  With the additional
approximation that $\Gamma_k\ll\omega_k$, we find
\begin{equation}
2(\delta n_k+\delta\nu_k)\approx-\frac{\partial_tn_k^{\mathrm{eq}}}{\Gamma_k^{\mathrm{eq}}}
+\frac{\dot{\omega}_k\Gamma_k^{\mathrm{eq}}(1+2n^{\mathrm{eq}})}{\omega_k^3},
\end{equation}
which reproduces the sum of (\ref{renums}) and (\ref{nkhs}).

It is not surprising that the above strategy agrees with linear response theory only
when $\Gamma_k\ll\omega_k$.  This indicates only that the methods available for resumming
self-energies in the equilibrium and non-equilibrium theories differ beyond the leading
order in $\Gamma_k/\omega_k$.  The fact that we can recover the linear response result only
by ignoring the fluctuations in self-energies induced by those in the $n_k$ is, however,
rather more significant, as we discover by attempting to solve the next-to-leading order
equations without this extra approximation. To simplify matters, we continue to retain
only the leading terms in $\Gamma_k/\omega_k$, in which case the order $\epsilon$ equations
are
\begin{equation}
{\mathrm{Re}}\,\delta\nu_k\approx\frac{1}{2\omega_k^2}
\left[-\omega_k\partial_t\left({\mathrm{Im}}\,\nu_k^{\mathrm{eq}}\right)
+\frac{\dot{\omega}_k\Gamma_k^{\mathrm{eq}}(1+2n_k^{\mathrm{eq}})}{2\omega_k}\right]
\label{deltanueqn}
\end{equation}
\begin{equation}
\int_0^\infty dk'K(k,k')\delta n_{k'}=\partial_tn^{\mathrm{eq}}_k-\frac{\dot{\omega}_k}{\omega_k}
{\mathrm{Re}}\,\nu_k^{\mathrm{eq}},
\label{deltaneqn}
\end{equation}
where
\begin{equation}
K(k,k') = \left.\frac{\delta S_k}{\delta n_{k'}}\right\vert_{n=n^{\mathrm{eq}}}.
\label{kernel}
\end{equation}
While the first of these gives $\delta\nu_k$ explicitly, the second is an integral
equation to be solved for $\delta n_k$.  It turns out that {\it this equation has no
solution}.  The reason is that the scattering processes described by $S_k$ conserve both
particle number and energy.  One easily finds that this implies the two sum rules
\begin{equation}
\int_0^\infty dk\,k^2K(k,k') = \int_0^\infty dk\,k^2\omega_kK(k,k')=0
\label{sumrules}
\end{equation}
valid for all $k'$.  The source terms on the right-hand side of (\ref{deltaneqn}) do not
respect these sum rules, so the equation is not self-consistent and has, in principle,
no solution.  At higher orders of perturbation theory, particle number is not conserved;
only the energy sum rule remains, but that is sufficient to invalidate (\ref{deltaneqn}).

It is important to emphasize that the evolution equation (\ref{dndt2}) with $\epsilon=1$
is perfectly sound:  it is a Boltzmann equation with a source term, which presumably has
a satisfactory solution for $n_k(t)$. What we have found is that this solution does not
have a time-derivative expansion.  That is, it cannot be expressed as
$n_k(t)=n_k^{(0)}(\phi)+n_k^{(1)}(\phi)\dot{\phi}+\ldots$ .  Nor, therefore, can the
equation of motion (\ref{phieom}).  Our principal conclusion, then, is that {\it the
friction coefficient} $\eta(\phi)$ {\it does not exist}.  As it stands, this conclusion
rests on an approximate treatment of a particular model, the $\lambda\Phi^4$ theory. It is
likely, however, to be quite generic, as we discuss in section \ref{discuss}.

\section{Numerical investigation\label{numeric}}
Although we have just reached the conclusion that the friction coefficient $\eta(\phi)$
does not exist, we have also seen that the linear-response result for $\eta(\phi)$ can
be recovered by ignoring fluctuations in the self-energies---an approximation that,
at first sight, would seem not to be severe for a system reasonably close to equilibrium.
Thus, although $\eta(\phi)$ formally does not exist, the local equation of motion
(\ref{localeom}) with $\eta(\phi)$ as given by linear response theory might be a reasonable
approximation to the true equation of motion.  We have obtained numerical results that
may bear on this question by taking advantage of the following circumstance.  A discretized
and truncated version of (\ref{deltaneqn}) that one might attempt to solve numerically is
\begin{equation}
\sum_{k'=0}^{k_{\mathrm{max}}}K_{k,k'}\delta n_{k'}=b_{k},
\label{deltantrunc}
\end{equation}
where $b_k$ stands for the source terms on the right-hand side.  Because the kernel $K_{k,k'}$
now involves only values of $k$ and $k'$ up to the cutoff value $k_{\mathrm{max}}$, it does
not exactly obey the sum-rule constraints (\ref{sumrules}) and the truncated equation may have
a solution.

In fact, we find that it has a very well defined solution, as illustrated in
\begin{figure}
\begin{center}
\includegraphics{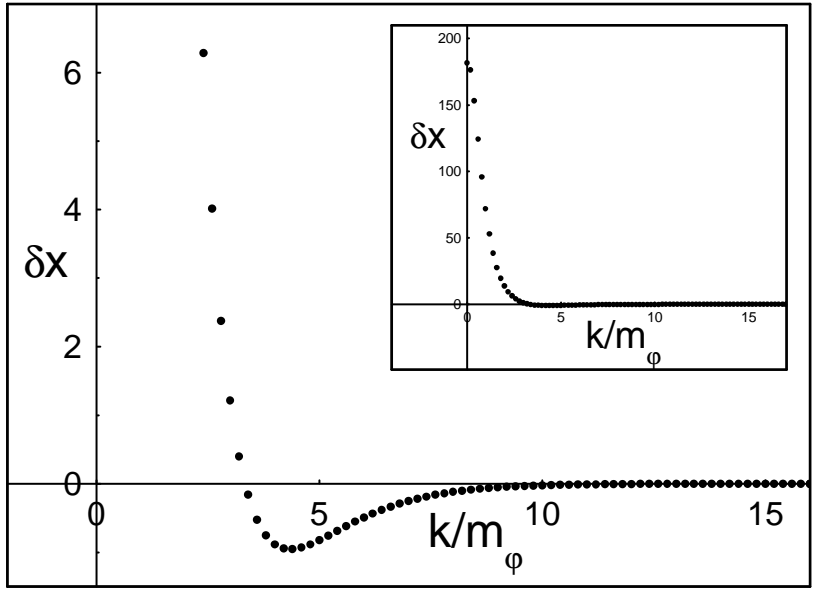}
\end{center}
\caption{\label{fig2}The solution of (\ref{deltantrunc}) for $\beta m_\varphi=1$ and $c=1$.}
\end{figure}
figure~\ref{fig2}, where the quantity plotted is
$\delta x = (4m^3c/\lambda\phi\dot{\phi})\delta n$ and we define the naturally-occurring
coupling constant $c = \lambda^2\pi^2/64(2\pi)^5$.  The example shown in figure~\ref{fig2}
is for $\beta m_\varphi=c=1$.  The kernel $K_{k,k'}$ decays rapidly for $k\gg k'$, but is
\begin{figure}
\begin{center}
\includegraphics{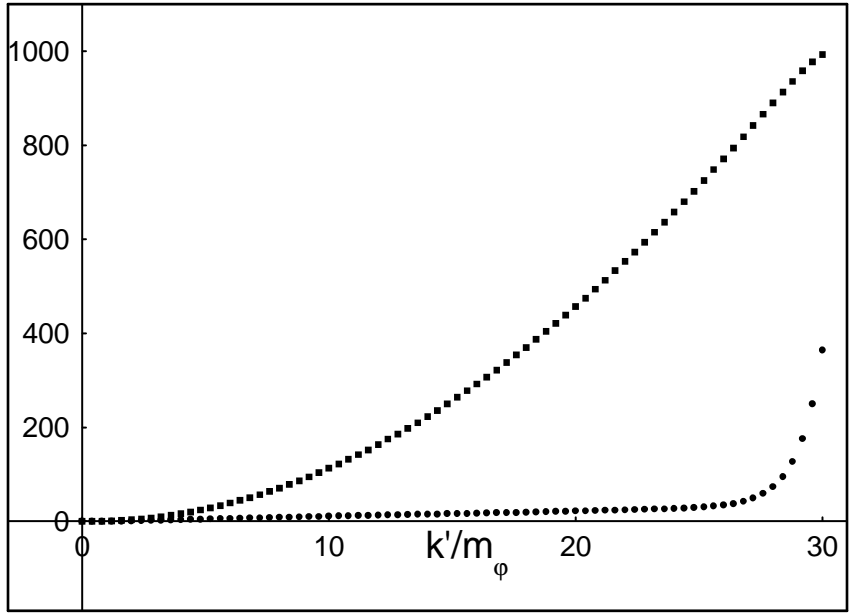}
\end{center}
\caption{\label{fig3}Illustration of the energy sum rule (\ref{sumrules});  the quantities
plotted are explained in the text.}
\end{figure}
not small when $k\lesssim k'$.  Thus, when both $k$ and $k'$ are bounded by the cutoff
$k_{\mathrm{max}}$, the sum rules may be verified for $k'\ll k_{\mathrm{max}}$, but they
fail for larger values of $k'$.  Figure~\ref{fig3} illustrates this for the energy sum rule, with
$\beta m_\varphi=1$ and a cutoff $k_{\mathrm{max}}=30m_\varphi$.  The circles in this figure
represent the integral $\int_0^{k_{\mathrm{max}}}dk\,k^2\omega_kK(k,k')$, while the squares
show, for comparison, the quantity $0.1\int_0^{k_{\mathrm{max}}}dk\,k^2\omega_k\vert K(k,k')\vert$.
The source $b_k$ in (\ref{deltantrunc}) becomes very small when $k$ is greater than a few times
$m_\varphi$, and figure 2 shows that the same is true of the solution $\delta n_k$.  In effect,
we see that, regardless of the cutoff, only a ``self-truncated'' kernel, with $k$ and $k'$
restricted to values smaller than a few times $m_\varphi$, contributes significantly to
the solution of (\ref{deltantrunc}).  The fact that this ``self-truncated'' kernel does not
in itself respect the sum rules accounts for the existence of a well defined solution and,
because of the self-truncation, we are able to verify that this solution converges to a
cutoff-independent form as $k_{\mathrm{max}}$ is increased.

\begin{figure}
\begin{center}
\includegraphics{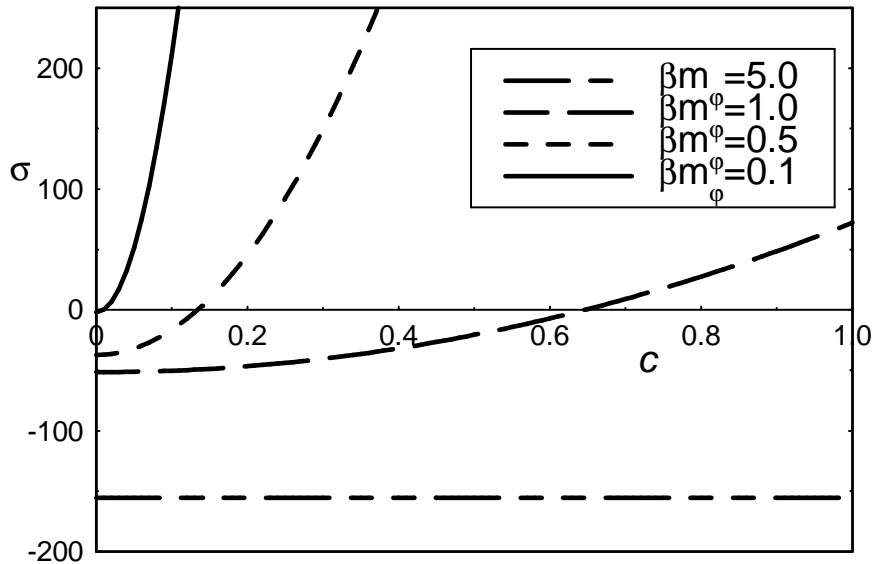}
\end{center}
\caption{\label{fig4}Friction coefficient calculated from the solution of (\ref{deltantrunc}).}
\end{figure}

\begin{figure}
\begin{center}
\includegraphics{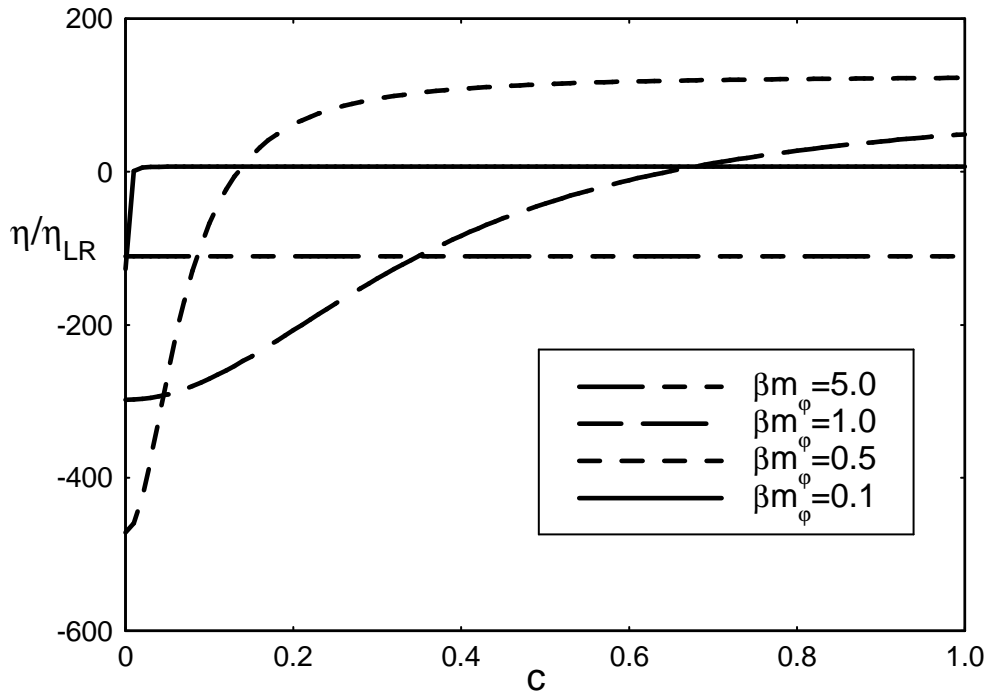}
\end{center}
\caption{\label{fig5}Ratio of the formal friction coefficient of figure~\ref{fig4} to that
calculated from linear response theory.}
\end{figure}

Formally, we can use this numerical solution in (\ref{approxev}) to obtain an estimate for
the friction coefficient $\eta$. The result of doing this for a range of coupling strengths
is shown in figure~\ref{fig4}, where the quantity plotted is
$\sigma = \eta m_\varphi/(128\pi\phi^2)$.
Figure~\ref{fig5} shows the ratio $\eta/\eta_{\mathrm{LR}}$, where $\eta_{\mathrm{LR}}$ is the
friction coefficient calculated in linear response theory from (\ref{etaI}) and (\ref{etaII})
in the limit $\Gamma_k\ll\omega_k$. (The haphazard appearance of this figure results from the
different dependencies of $\eta_{\mathrm{I}}$ and $\eta_{\mathrm{II}}$ on temperature and
coupling strength.)
The formal quantity shown in figure~\ref{fig4} has negative values at
weak coupling (where our perturbative methods are most likely to make sense), and clearly
cannot be interpreted as a genuine friction coefficient.  This, of course, is consistent
with our earlier argument that the friction coefficient is not well defined.  The conclusion
is that non-equilibrium methods are needed to investigate the real effect of dissipation
in the equation of motion (\ref{geneom}), even for a system that may be quite close to
equilibrium.  One approximate method of doing this is to integrate this equation of motion
simultaneously with the evolution equations (\ref{dndt}) and (\ref{dnudt})
and we plan to report on such calculations in future work.  The large discrepancies
apparent in figure~\ref{fig5} suggest that quantitatively significant deviations from
the predictions of linear response theory may be expected.

\section{Friction arising from fermions\label{fermions}}
We comment briefly on the frictional effect of a Yukawa coupling to fermions.  Non-equilibrium
perturbation theory for fermions is discussed in Ref.~\cite{lawrie2000}, to which we refer
the reader for the somewhat cumbersome details.  With approximations analogous to those
described above for scalar fields, we find for the relevant term in (\ref{geneom})
\begin{equation}
\langle\bar{\psi}(t,\bm{x})\psi(t,\bm{x})\rangle = 4\int\frac{d^3k}{(2\pi)^32\omega^{(\psi)}_k}
\left[m_\psi(2n_k^{(\psi)}-1)+2\vert\bm{k}\vert{\mathrm{Re}}\,\nu_k^{(\psi)}\right]
\label{fermionev}
\end{equation}
where $n^{(\psi)}_k(t)$ and $\nu^{(\psi)}_k(t)$ are the fermionic analogues of the functions
$n_k(t)$ and $\nu_k(t)$.  If the fermions have a mass $m_\psi(t)<\frac{1}{2}m_\varphi(t)$,
then the on-shell processes $\varphi\rightleftharpoons \bar{\psi}\psi$ are kinematically
allowed.  We then obtain kinetic equations of the form
\begin{eqnarray}
\partial_tn_k^{(\psi)}&=&-\Gamma_k^{(\psi)}\left(2n_k^{(\psi)}-1\right)
+\frac{\omega_k^{(\psi)}}{\vert\bm{k}\vert}\alpha_k^{(\psi)}
+\frac{\dot{\omega}_k^{(\psi)}}{\omega_k^{(\psi)}}\frac{\vert\bm{k}\vert}{m_\psi}
{\mathrm{Re}}\,\nu_k^{(\psi)}\label{dndtf}\\
\partial_t\nu_k^{(\psi)}&=&-2i\left(\omega_k^{(\psi)}-i\Gamma_k^{(\psi)}\right)
\nu_k^{(\psi)}-\left(\frac{\omega_k^{(\psi)}-m_\psi}{\omega_k^{(\psi)}}\right)
\alpha_k^{(\psi)}-\frac{\dot{\omega}_k^{(\psi)}}{2\omega_k^{(\psi)}}
\frac{\vert\bm{k}\vert}{m_\psi}\left(2n_k^{(\psi)}-1\right)\label{dnudtf}
\end{eqnarray}
where  $\alpha_k^{(\psi)}$ and $\Gamma_k^{(\psi)}$ are extracted as in (\ref{alphapresc})
and (\ref{gammapresc}) from the fermion self-energy.  Naturally, the scalar self-energy
now acquires a contribution from a fermion loop.  Taking this into account, the kinetic
equations (\ref{dndt}) and (\ref{dndtf}) are a consistent pair of Boltzmann equations,
in which the scattering integrals preserve the particle number $2N_\varphi+N_\psi$ (we have
derived (\ref{dndtf}) and (\ref{dnudtf}) only for zero chemical potential, in which case
fermions and antifermions are equally abundant and $N_\psi$ means the total number of these
particles) and the total energy of scalar and fermionic particles.  These conservation laws
(or, at higher orders, just energy conservation) again imply that neither the solution of
the Boltzmann equations nor the equation of motion for $\phi$ has a time-derivative expansion.

\section{Discussion\label{discuss}}
The equation of motion (\ref{geneom}) is inherently non-local in time;  it represents
a non-Markovian process in which evolution depends on the history $\phi(t')$ at all times
prior to the time $t$ of interest.  When the state of the system is not too far from thermal
equilibrium, it is tempting to suppose that a local equation of motion equation (\ref{localeom})
might be approximately valid, the friction coefficient being estimated from equilibrium
statistical mechanics.  In this paper, we have examined the approximations needed to
extract a local equation of motion from the non-local one, and concluded that this cannot
in fact be done.  Under suitable conditions (the principal requirement is the existence of a
relaxation time $\Gamma_k^{-1}$ short enough to ensure that correlations decay rapidly compared
with the characteristic time scale on which $\phi(t)$ changes) the expectation values in
(\ref{geneom}) can be approximated by local expressions of the form (\ref{approxev}) or
(\ref{fermionev}), in which the auxiliary functions $n_k(t)$ and $\nu_k(t)$ themselves obey
local kinetic equations, such as (\ref{dndt}), (\ref{dnudt}), (\ref{dndtf}) and (\ref{dnudtf}).
However, this set of local evolution equations can be reduced further to a single local equation
for $\phi(t)$ only if the kinetic equations admit a solution in the form of a time-derivative
expansion---and we find that they do not.  The obstruction arises from fluctuations in
self-energies, of which the equilibrium theory takes no account.

Although our explicit computations focussed on the simplest example of a single, self-coupled
scalar field, we have indicated is sections \ref{intro} and \ref{fermions} that the situation
is quite generic.  The above conclusion emerges from an approximate treatment of non-equilibrium
dynamics, which can hardly be regarded as a rigorous proof.  It would seem that some approximation
more or less equivalent to the local {\it ans\"atze} (\ref{approxsigmarho}) and (\ref{approxsigmac})
for self-energies is an inevitable step towards the derivation of a local equation of motion; without
some such approximation, the expectation values in (\ref{geneom}) remain non-Markovian and,
{\it a fortiori}, cannot be represented by a local friction term. A subsidiary approximation
made in section~\ref{friction} was to set the quasiparticle width $\Gamma_k$ to zero for the
purposes of estimating self-energies.  Although this approximation greatly simplified our
analytical analysis, it can and should be avoided in a comprehensive numerical study.  The effect
of this approximation is to restrict the scattering processes in Boltzmann equations to on-shell
processes.  Now, the inclusion of off-shell processes might well invalidate the sum rules
(\ref{sumrules}) from which we concluded that the friction coefficient $\eta$ does not
exist.  Formally, then, by including off-shell processes, we might after all be able to obtain a
time-derivative expansion of the equation of motion.  However, the friction coefficient implied
by this expansion would be quantitatively similar to that obtained in section~\ref{numeric}.  At
weak coupling, it is negative (and thus physically unacceptable) and quite different from the
one yielded by linear response theory.  At strong coupling, the perturbative methods employed
both here and in linear response theory are quantitatively, and perhaps also qualitatively,
unreliable.  Our practical conclusion, then, is that even for a system quite close to thermal
equilibrium the local equation of motion (\ref{localeom}) does not furnish an reliable
account of dissipation, whether or not off-shell processes serve to recover a formal time-derivative
expansion.  A thorough numerical investigation of the non-equilibrium evolution is therefore
essential.  At weak coupling, a numerical implementation of the evolution equations developed here
is quite widely applicable, and may well be quantitatively adequate, though recently developed methods
based on the 2PI-1/N formalism \cite{aarts,berges2002} are probably more powerful in situations where
they can be applied.

Finally, we observe that the analysis given here applies to quantum field theory in an expanding
spacetime with only minor modifications.  In a spatially homogeneous Robertson-Walker spacetime,
the field redefinitions $\Phi\to a^{-1}\Phi$ and $\psi\to a^{-3/2}\psi$, where $a(t)$ is the scale
factor, serve to cast the theories we deal with in the form of a Minkowski-space theory with
time-dependent masses, provided that $t$ is taken as the conformal time coordinate.  In the case of
a spinor field or a conformally coupled scalar field, these masses are given simply by $m(t)=a(t)m$.
Consequently, the formalism we have constructed changes only insofar as the masses in (\ref{masses})
depend on $t$ through both $\phi(t)$ and $a(t)$.  This additional time dependence modifies evolution
equations in a way that may be cosmologically important.  However, its net effect on, say, equations
(\ref{deltanueqn}) and (\ref{deltaneqn}) is just that the right-hand sides of these equation
contain terms proportional to $\dot{a}$ in addition to those proportional to $\dot{\phi}$.  These induce
additional contributions to $\delta n_k$ and $\delta\nu_k$, but do not affect our conclusions concerning
the terms proportional to $\dot{\phi}$.

\begin{acknowledgments}
It is a pleasure to acknowledge informative discussions with Arjun Berera, Rudnei Ramos
and Jim Morgan.
\end{acknowledgments}

\end{document}